\documentclass[fleqn,usenatbib]{mnras}
\usepackage{hyperref}
\usepackage{newtxtext,newtxmath}

\usepackage[T1]{fontenc}
\usepackage{ae,aecompl}

\usepackage{graphicx}	
\usepackage{amsmath}	
\usepackage{amssymb}	

\newcommand{\markus}[1]{{\color{black} #1}}
\newcommand{\sk}[1]{{\color{black} #1}}

\title[Calibrating LPV as standard candles with ML]{Calibrating Long Period Variables as Standard Candles with Machine Learning}

\author[]{
Markus Michael Rau$^{1}$\thanks{E-mail: markusr@andrew.cmu.edu},
Sergey E. Koposov$^{1,2}$,
Hy Trac$^{1}$,
Rachel Mandelbaum$^{1}$
\\
$^{1}$McWilliams Center for Cosmology, Department of Physics, Carnegie Mellon University, Pittsburgh, PA 15213\\
$^{2}$ Institute of Astronomy, University of Cambridge, Madingley Road, Cambridge CB3 0HA, UK
}

\date{Accepted XXX. Received YYY; in original form ZZZ}

\pubyear{2015}

\begin{document}
\label{firstpage}
\pagerange{\pageref{firstpage}--\pageref{lastpage}}
\maketitle

\begin{abstract}
Variable stars \markus{with well-calibrated period-luminosity relationships} provide accurate distance measurements to nearby galaxies and are therefore a vital tool for cosmology and astrophysics. While these measurements typically \markus{rely on} samples of Cepheid and RR-Lyrae stars, abundant populations of luminous variable stars with longer periods of $10 - 1000$ days remain largely unused. We apply machine learning to derive a mapping between lightcurve features of these variable stars and their magnitude to extend the traditional period-luminosity (PL) relation commonly used for Cepheid samples. Using photometric data for long period variable stars in the Large Magellanic cloud (LMC), we demonstrate that our predictions produce residual errors comparable to those obtained on the corresponding Cepheid population. We show that our model generalizes well to other samples by performing a blind test on photometric data from the Small Magellanic Cloud (SMC). Our predictions on the SMC again show small residual errors and biases, comparable to results that employ PL relations fitted on Cepheid samples. The residual biases are complementary between the long period variable and Cepheid fits, which provides exciting prospects to better control sources of systematic error in cosmological distance measurements. We finally show that the proposed methodology can be used to optimize samples of variable stars as standard candles independent of any prior variable star classification.
\end{abstract}




\section{Introduction}
One of the most important aspects of modern cosmology and astrophysics is the measurement of accurate distances. Variable stars like Cepheids that exhibit tight relationships between their oscillation period and luminosity are among the primary tools to measure distances in the local universe. In the advent of precision cosmology, Cepheid distances are a vital rung in the cosmological distance ladder and calibrate local Type Ia supernovae (SNIa) samples. These SNIa distance measurements \citep[e.g.][]{2011ApJS..192....1C, 2016ApJ...826...56R} provide an absolute distance scale in the low redshift universe, that complement constraints on the CMB sound horizon scale \citep{2016A&A...594A..13P} at the high redshift border of the visible universe. The mild tension between both distance scales that currently persists in the literature \citep[e.g.][]{2016ApJ...826...56R,  2017MNRAS.471.2254Z, 2018MNRAS.476.3861F} could \markus{require new theoretical interpretations of these} complementary distance scales. Given sufficient \markus{observational} evidence, this can motivate considering non-standard extensions to the cosmological model like the introduction of sterile neutrinos, dynamical dark energy or a nonzero curvature component \citep[e.g.][]{PhysRevLett.112.051302, 2014PhRvD..90h3503D, 2014PhRvL.113d1301L, 2016A&A...594A..14P, 2016PhLB..761..242D, 2016JCAP...10..019B, 2017NatAs...1..627Z, SOLA2017317}. A primary challenge in the field is therefore to clarify if the observed tensions are significant signs of new physics or caused by observational systematics. 

Cepheid distances, which provide the primary calibration for local supernovae samples, are subject to a variety of potential observational and methodological systematics \citep[e.g.][]{2001ApJ...553...47F, 2011ApJS..192....1C, 2014MNRAS.440.1138E, 2015ApJ...799..144K, 2016ApJ...826...56R, 2017MNRAS.471.2254Z, 2018MNRAS.476.3861F} such as inaccurate photometric calibration, metallicity differences between anchor samples or biases introduced by the treatment of outliers in fits of the period-luminosity (PL) relation. In addition, sufficiently large anchor samples for Cepheids are only available in $\sim 9$ nearby galaxies \citep{2017MNRAS.471.2254Z}. As a result, $\approx 30\%$ of the total error budget on local $H_0$ measurements is related to the Cepheid distance calibration \citep[see][Fig. 1]{2016ApJ...826...56R}. Exploring additional sources of distance calibration for local supernovae measurements is therefore an interesting avenue to better control sources of systematic errors in $H_0$ measurements. For example \citet{2018arXiv180102711H} recently proposed the tight period luminosity relation of oxygen-rich Mira variables as an additional rung in the cosmological distance ladder. Distance measurements that use variable stars exploit the tight relationship between period, metallicity and the luminosity of Cepheids, RR-Lyrae and Mira variables to calibrate them as cosmological standard candles. However, there exist a large variety of other luminous variable stars, like OGLE Small Amplitude Red Giant stars (OSARG), that also exhibit a variety of overlapping sequences in PL space.

The goal of this paper is to investigate how machine learning can be used to exploit these variable stars for cosmological distance measurements. This is facilitated by the rich feature set in variable star lightcurves, typically used in the context of variable star classification \citep[e.g.][]{2011ApJ...733...10R, 2011MNRAS.414.2602D, 2013AJ....146..101P, 2015MNRAS.451.3385K, 2016MNRAS.456.2260A, 2017AJ....153..204S, 2018NatAs...2..151N}, that contains much more information about the variable star luminosity than just the first dominant period. 

This paper is structured as follows. In \S \ref{sec:data} we will describe the photometric data and the lightcurve features used in this analysis. \S \ref{sec:methodology} describes our methodology, the metrics used, and additional analysis steps such as outlier rejection. \S\ref{sec:analysis_results} details our analysis and \S \ref{sec:conclus} closes with a discussion and summary of our results.

\section{Data}
\label{sec:data}
We use star catalogs from the third public release of the Optical Gravitational Lensing Experiment (OGLE) survey \citep[e.g.][]{ogle_III_database, 2008AcA....58..163S, 2009AcA....59..239S, 2010AcA....60...17S, 2011AcA....61..217S}, which provides photometric lightcurves of the Large Magellanic Cloud and Small Magellanic Cloud in the $I$ and $V$ bands.  The catalog contains variable star classifications, mean lightcurve magnitudes in the $I$ and $V$ filters and the three primary periods $P_{1-3}$ and amplitudes $A_{1-3}$ extracted from the $I$ band lightcurves by Fourier decomposition \markus{as described in \citet{2004AcA....54..129S}.} We use the fundamental mode Cepheid and the Long Period variables (LPV) in the full catalog, where the LPV sample consist of semi-regular variables (SRV), Mira stars and OGLE Small Amplitude Red Giant (OSARG) variables. \markus{In the following we will further distinguish stars on the Red Giant Branch (RGB) and the Asymptotic Giant Branch (AGB).} We correct the $I$ \markus{band} magnitude for reddening using the optical Wesenheit index \citep[e.g.][]{1982ApJ...253..575M, 1998ApJ...500..525S, 2007AcA....57..201S} 
\begin{equation}
 W_{I} = I - 1.55 \cdot (V - I) \, , 
 \label{eq:wesenheit_eq}
\end{equation}
where $I$ and $V$ denote the mean apparent magnitudes in the $I$ and $V$ band respectively. \markus{We restrict the Wesenheit range of the LPV sample to $3<W_I<16$ and use $|W_{I}| < 17.5$ for the fundamental mode Cepheid sample, which is slightly fainter than the LPV population. Beyond these Wesenheit magnitude limits we only find LPV and fundamental mode Cepheid stars that lack magnitude measurements in the $V$ or $I$ band (values are set to -99.99) or a total of 14 (7) very faint LPV in the LMC (SMC) with up to $W_I \lesssim 20$.}
\begin{figure*}
   \centering  
   \includegraphics[width=0.95\textwidth]{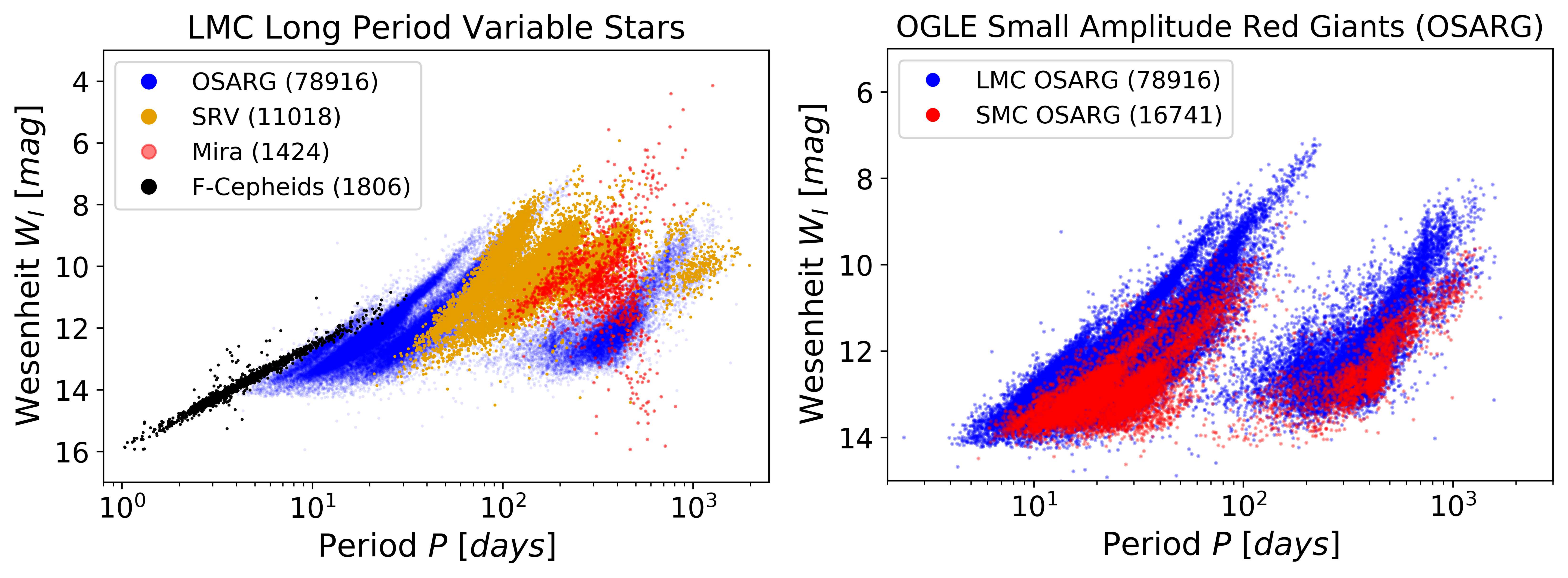}
   \caption{\label{fig:overview_p_w} Overview of the Period-Wesenheit (P-W) relations for the different subsamples used in this analysis. The legends list the number of stars in the respective subsamples in brackets. \textit{Left}: Fundamental Mode Cepheids (F-Ceph), Mira stars (Mira), semi-regular variables (SRV) and OGLE Small Amplitude \markus{Red Giant stars} (OSARGs) in the Large Magellanic Cloud (LMC). \textit{Right}: P-W relation of OSARG variables in the LMC and the Small Magellanic cloud (SMC).  } 
\end{figure*}
\begin{figure*}
   \centering  
   \includegraphics[scale=0.6]{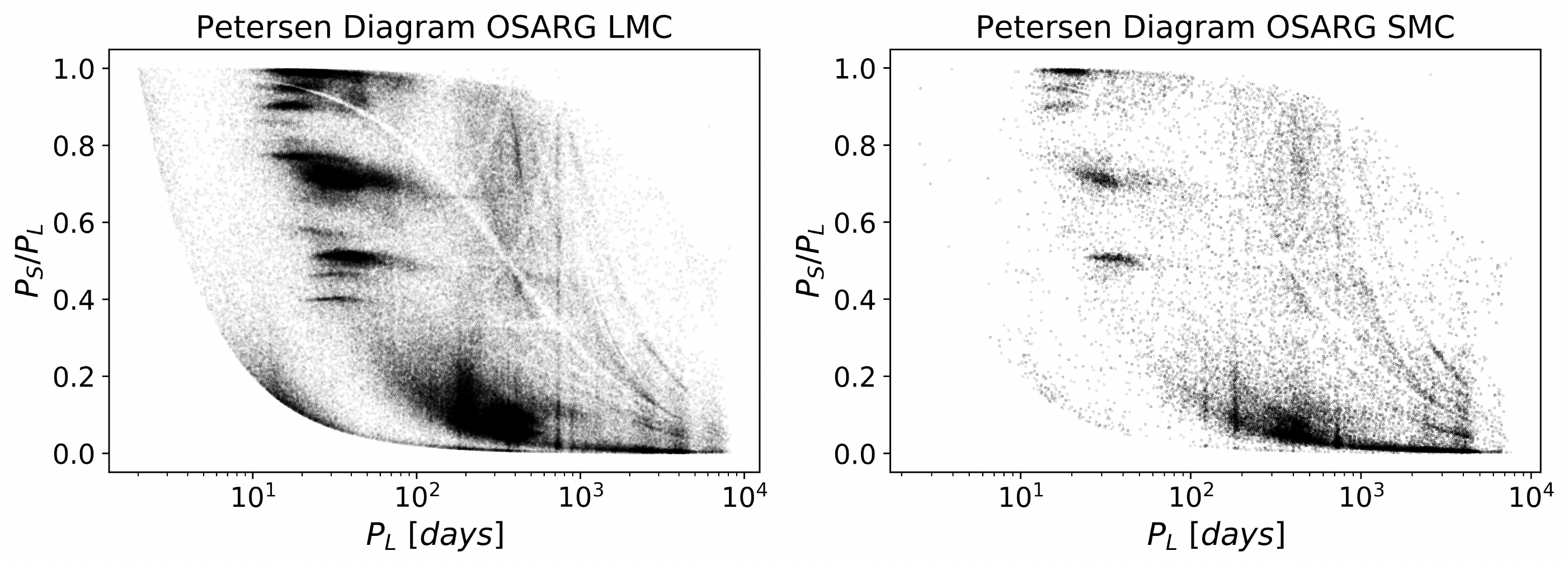}
   \caption{\label{fig:petersen_diagram} \markus{Petersen diagram of the full OSARG AGB and RGB star samples in the LMC (left panel) and SMC (right panel).} Forming all combinations from the three primary periods, we plot the longer period $P_L$ against the ratio between shorter and longer period $P_S/P_L$. \markus{A star can therefore appear up to 3 times in the Petersen diagram, depending on the number of measured periods.}} 
\end{figure*}

Fig.~\ref{fig:overview_p_w} plots the first period against the optical Wesenheit index (or `P-W relation'), for the different types of variable stars in the selected catalog. The left panel shows the P-W relations for the population in the Large Magellanic Cloud (LMC). Besides the population of fundamental mode Cepheids (F-Cepheids), we see a large sample of bright variable stars with longer periods covering two orders of magnitude from $10$ to $1000$ days. The sample size of each stellar population is shown in the legend. The largest population, $\approx 80.000$ variable stars, are OGLE Small Amplitude Red Giant stars (OSARGs). Notably, the OSARG sample is significantly brighter than the corresponding Cepheid sample for $P > 20$ days. The right panel of Fig.~\ref{fig:overview_p_w} shows the P-W relation of OSARGs in the LMC (blue) and SMC (red). Besides the shift towards fainter magnitudes due to the distance modulus between the two galaxies, we also note that the stellar population is significantly different. While both LMC and SMC OSARGs follow the same characteristic P-W \markus{sequences}, the SMC OSARGs do not cover the full range of magnitudes and periods that would be expected from the LMC population. Furthermore the P-W relations in the SMC appear slightly tilted compared with the LMC \markus{sequences}. A likely explanation for this discrepancy could be differences in the metallicity between both stellar populations. 
\begin{figure*}
   \centering  
   \includegraphics[width=0.95\textwidth]{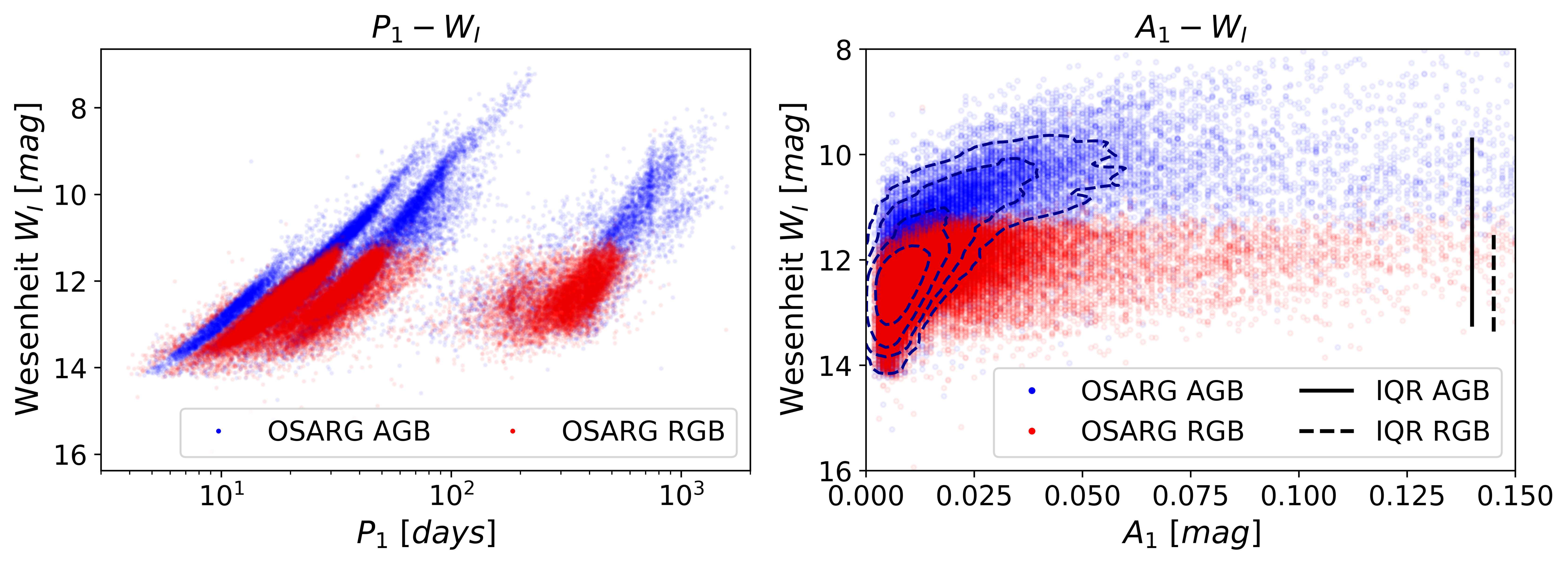}
   \caption{\label{fig:features} Differences in the primary period-amplitude distribution between the OSARG AGB and OSARG RGB subsamples. \textit{Left}: Primary period $P_1$ against Wesenheit $W_I$ distribution for the AGB (blue) and RGB (red) subsamples. \textit{Right}: Primary amplitude $A_1$ against Wesenheit $W_I$ distribution for the AGB (blue) and RGB (red) subsample. The vertical lines show the inter quantile range of the $W_I$ distribution between the 0.05 and 0.95 quantiles of the respective subsamples. \markus{The blue dashed density contours highlight the $A_1$ against $W_I$ distribution for OSARG AGB variables, that is partly overplotted by the OSARG RGB points.}} 
\end{figure*}

\subsection{Features: Periods}
\label{subsec:periods}
The overlapping P-W sequences {\color{black} that are in the current literature referred to as} OSARG, Mira and SRV \markus{stars} were initially described by \citet{1999IAUS..191..151W} and will be referred to as `Wood sequences'. They correspond to modes of stellar oscillation and can be related to clusters of period ratios. These can be efficiently analysed using so-called Petersen diagrams, shown in Fig.~\ref{fig:petersen_diagram}. This diagram, {\color{black} applied earlier by e.g. \citet{2004AcA....54..129S}}, shows the distribution of pulsators in the space of longest period vs.\ shorter period to long period ratio. The left (right) panel shows the Petersen diagram for the OSARG population in the LMC (SMC). \citet{2004AcA....54..129S} studied the OSARG population in the Magellanic Clouds and showed that the clusters in the Petersen diagram can be associated with the Wood sequences and thus to the type of stellar pulsation mode. Period ratios around 0.97 to 1.0 can be associated with non-radial oscillations that can also be found in \markus{Cepheids\footnote{We refer to \citet{2017EPJWC.15206003S} for a more detailed discussion of Petersen diagrams for Cepheids and RR-Lyr.} and RR-Lyr \citep[][and references therein]{2004AcA....54..129S}}. 

The lower right clump around $P_L \approx 200$ days can be associated with Long Secondary Period (LSP) {\color{black} oscillations}, whose origin is still an area of active research \citep[see][and references therein]{2017ApJ...847..139T}. The near horizontal sequences visible in the Petersen diagram are related to the aforementioned Wood sequences in P-W space and thus correspond to characteristic period ratios of stellar oscillations. Comparing the left and right panels of Fig.~\ref{fig:petersen_diagram}, we see that the basic structure of the Petersen diagram is the same in both the LMC and SMC samples. While the number of OSARG stars is lower in the SMC than in the LMC, the {\color{black} horizontal clusters of stars with similar period ratios at $P_L < 10^2$} are populated in both diagrams. We refer to \citet{2004AcA....54..129S} for a more detailed analysis of Petersen diagrams of OSARG variables in the Magellanic Clouds. 

The left panel of Fig.~\ref{fig:features} shows the P-W relation for the RGB and AGB populations of LMC OSARGs. As expected, the AGB stars extend towards brighter magnitudes, but largely cover the same Wood sequences as the RGB population, in agreement with \citet{2003MNRAS.343L..79K, 2004AcA....54..129S}. We note a small period shift between the RGB and AGB samples, which can be explained by the fact  that the characteristic oscillation period $P$ for solar oscillations scales with effective temperature $T_{\rm eff}$, stellar mass $M$ and luminosity $L$ as $P \sim L / (M \, T_{\rm eff}^{3.5})$ as given in \citet{2007MNRAS.377..584S}. At constant luminosity the AGB stars will have a higher effective temperature compared to RGB stars. This temperature difference induces a small period shift between samples of AGB and RGB stars \citep{2003MNRAS.343L..79K, 2004AcA....54..129S}. 

We conclude that the first three dominant periods contain information about the type of stellar oscillation, the position on the respective Wood sequences as well as the evolutionary state of the giant star. {\color{black} The luminosity information contained in the multiple oscillation periods of RGB stars was already exploited for distance measurements by e.g. \citet{2010MNRAS.409..777T}. Accordingly the multiple oscillation periods can be expected to be important features in our Machine Learning approach.} 

\subsection{Features: Amplitudes}
\label{subsec:amplitudes}
The right panel of Fig.~\ref{fig:features} plots the first amplitude $A_{1}$ against the Wesenheit magnitude $W_{I}$ for the AGB and RGB OSARG variables. We plot the range between the 0.05 and 0.95 quantile, i.e. the inter-quantile range ${\rm IQR}$, for the AGB (RGB) population as solid (dashed) vertical lines and highlight the amplitude-Wesenheit distribution of the OSARG AGB population with density contours. The amplitude and the Wesenheit are \markus{anti-correlated}, where AGB and RGB OSARG samples with smaller amplitudes extend towards fainter magnitudes as to be expected from the peak amplitude scaling $A \sim L / (M \, T_{\rm eff}^2)$ predicted by \citet{2007MNRAS.377..584S}. As discussed in \citet{2017ApJ...847..139T}, the observed amplitudes are related to the growth rate of the stellar oscillation modes and can therefore be expected to help in distinguishing between different modes of stellar oscillation. 

From these considerations we conclude that the inclusion of the first three amplitudes into the feature set is  well motivated by both their correlation with the luminosity of the star and by their connection to the growth rates of the respective stellar oscillation modes.

\section{Methodology}
\label{sec:methodology}
As noted previously, our goal is to select samples of long period variables for which we can obtain accurate Wesenheit predictions. In this way we are able to jointly optimize the sample selection, i.e.\ the identification of `standard candle-like' stars, and make accurate predictions of the Wesenheit magnitudes based on the high dimensional feature set. The following analysis uses the lightcurve features discussed in \S \ref{subsec:periods} and \ref{subsec:amplitudes}: 
\begin{itemize}
\item the primary $P_1$, secondary $P_2$ and tertiary $P_3$ oscillation periods; 
\item the corresponding amplitudes $A_1$, $A_2$ and $A_3$ of the oscillations.
\end{itemize}

The full information of the mapping between the lightcurve features $\mathbf{f}$ and the apparent Wesenheit index $W_{I}$ is contained in the conditional probability density function (pdf) $p(W_{I} | \mathbf{f})$. Once this distribution is estimated for each variable star in the catalog, we define a statistic of this distribution that will be used 
to obtain Wesenheit predictions for a given variable star. In the classical regression setting this is the conditional mean, which is also our choice, but alternatives such as the conditional median can be justified if $p(W_{I} | \mathbf{f})$ is expected to exhibit wide wings, due to a significant fraction of outliers in the data. The estimated distribution $p(W_{I} | \mathbf{f})$ is a convolution of a pdf that describes the photometric error in $W_{I}$, the intrinsic error of the data\footnote{\markus{Here, the intrinsic error refers to the standard deviation of the conditional pdf obtained by a perfect estimator on noiseless data. This error depends only on the intrinsic information in the lightcurve features to predict the Wesenheit, but not on inaccuracies in the estimator or data.}} and other effects like attenuation bias from inconsistencies in the errors on input features across different datasets \markus{or inaccuracies in the machine learning algorithm}. Since the OGLE photometry is of exceptional quality ($S/N > 1500$) and very similar in both the LMC and SMC, we do not incorporate the photometric error or attenuation bias into the modeling and assume that $p(W_{I} | \mathbf{f})$ is dominated by the intrinsic error in the data. 

As a selection criterion \sk {of high-quality standard candles} we use the standard deviation $\sigma(W_{I} | \mathbf{f})$ of the conditional pdf $p(W_{I} | \mathbf{f})$ and select only those objects for which this quantity is small. This is motivated by our previous decision to use the mean of $p(W_{I} | \mathbf{f})$ as the regression statistic. The procedure therefore essentially approximates the conditional pdf $p(W_{I} | \mathbf{f})$ as a normal distribution. The mean is used to predict the Wesenheit magnitude of the variable star and the standard deviation allows us to select those stars that are expected to occupy regions in feature space where the most accurate predictions are possible.
While the analysis \sk{presented in this paper} only relies on estimates for two statistics of the conditional pdf, \sk{the general method that we demonstrate} constructs the full shape of this distribution. This enables a possible extension to alternative point statistics like the conditional median.

\subsection{Conditional Density Estimation}
\label{subsec:cond_dens}
In this section we describe our machine learning methodology to construct an estimate for the conditional pdf $p(W_{I} | \mathbf{f})$ from which the mean and standard deviation are derived. To avoid overfitting, we split the available data randomly into two disjunct subsamples; the training and test set.
The model is then fitted, or trained, on the training set and subsequently applied to the disjunct test set. \markus{This work will use 90\% of the data to train the model and 10\% as test data, in a so-called `10-fold cross validation' procedure described in \S \ref{subsec:evaluation_metrics}.}

To estimate the conditional pdf $p(W_{I} | \mathbf{f})$, we discretize the Wesenheit index of the training set into 300 equally spaced bins.\footnote{The results are not very sensitive to this choice, but choosing an overly-coarse binning can lead to biased estimates.} In this way we reformulate the regression problem of predicting the continuous Wesenheit index $W_I$ as a classification problem \citep[e.g.][]{ICML02-tac, DBLP:conf/acml/FrankB09, 2015MNRAS.452.3710R}. The model will then estimate probabilities of Wesenheit-bin membership. These probabilities can then be combined into a histogram that is an estimate of the conditional pdf $p(W_{I} | \mathbf{f}_i)$ for each variable star $i$ in the sample.

It is convenient to express this estimate as a weighted sum over the Wesenheit magnitudes $W_I^i$ of the stars $i$ in the training set. The Wesenheit interval $j \in [1, n_{\rm bins} = 300]$ into which the star $i$ falls is denoted as $\mathcal{I}_j$. Denoting the bin probability of bin $j$ as \markus{$\mathcal{P}_j$}, we can define a weight $w_i(\mathbf{f})$ for each variable star in the training set as \markus{
\begin{equation}
w_i(\mathbf{f}) = \sum_{j = 1}^{n_{\rm bins}} \left(\frac{\mathcal{P}_j}{n_j}\right)  \, \Theta(W_I^{i} \in \mathcal{I}_{j}) ,
\end{equation}}
where $n_j$ is the number of all variable stars in the training set that fall into bin $\mathcal{I}_j$ and $W_I^{i}$ denotes the Wesenheit magnitude of the training set \markus{star}. 
For a variable star in the test set with Wesenheit magnitude $W_I$ and feature vector $\mathbf{f}$, we can write the conditional distribution $p(W_I | \mathbf{f})$ as 
\begin{equation}
p(W_I | \mathbf{f}) = \sum_{i = 1}^{N} w_i(\mathbf{f}) \sum_{j = 1}^{n_{\rm bins}} \frac{\Theta(W_I \in \mathcal{I}_j) \, \Theta(W_I^{i} \in \mathcal{I}_j)}{r_j} \, ,
\label{eq:histogram_cde} 
\end{equation} 
where $N = \sum_{j = 1}^{j = n_{\rm bins}} n_j$ denotes the number of variable stars in the training set {\color{black} and} $r_j$ is the width of bin $\mathcal{I}_j$. {\color{black} The boolean function $\Theta(x)$ is 0 if its argument is false and unity if it is true. Accordingly} $\Theta(W_I \in \mathcal{I}_{j}) \, \Theta(W_I^{i} \in \mathcal{I}_{j})$ is unity if both $W_I^{i}$ and $W_I$ are in bin $\mathcal{I}_{j}$ and 0 otherwise. Note that these weights are a function of the features $\mathbf{f}$ of the respective variable star in the test set, as the bin probabilities will depend on the position in feature space. 

\markus{The conditional mean $\left\langle W_I | \mathbf{f}\right\rangle$ can be estimated on the weighted training set as
\begin{equation}
\left\langle W_I | \mathbf{f}\right\rangle = \sum_{i = 1}^{N} w_i(\mathbf{f}) \, W_I^{i} \, ,
\label{eq:cond_mean}
\end{equation}
and the conditional standard deviation $\sigma(W_I | \mathbf{f})$ as
\begin{equation}
\sigma(W_I | \mathbf{f}) = \sqrt{\sum_{i = 1}^{N} w_i(\mathbf{f}) \, \left(W_I^{i} - \left\langle W_I | \mathbf{f}\right\rangle\right)^2} \, 
\label{eq:cond_std}
\end{equation}}

In \S \ref{sec:analysis_results}, we rank the variable stars in order of increasing conditional standard deviation and select a subsample that contains only stars that are expected to yield very accurate predictions of their Wesenheit magnitude given their feature vectors $\mathbf{f}$. 

\subsection{The Random Forest}
In the following we describe the Random Forest\footnote{We use the implementation provided by the scikit-learn package \citep{scikit-learn}, using the default parameters from the `Random Forest Classifier'.} classifier \citep{Breiman:2001:RF:570181.570182} that we use to estimate the bin probabilities $\mathcal{P}_j$. 

Given a training set of variable stars with known combinations of features $\mathbf{f}_i$ and Wesenheit indices $W_{I}^i$, the algorithm starts by forming $N_T$ number of bootstrap realizations of this training set. In the process of bootstrapping, we randomly select $N$ elements from the original dataset with replacement, where $N$ denotes the sample size of the original dataset. On each of these bootstrap realizations a single decision tree is fitted. The index $T$ identifies a particular tree in the Random Forest. The Random Forest predicts bin probabilities by averaging the bin probabilities estimated by all $N_T$ decision trees in the Random Forest.  

A single decision tree is a binary partitioning tree that is recursively grown on the bootstrapped dataset by selecting splits such that the newly formed partitions are optimized to contain only training set stars of high similarity. Each grown partition with index $\tau \in [1, N_{\tau}]$ corresponds to a region in input space $\mathcal{R}_{\tau}$. We denote $N_{\tau}$ as the total number of partitions, or `leaf nodes' of the tree, and $n_{\tau}$ as the number of training set elements in partition $\tau$. The probability that a variable star whose features fall into a region $\mathbf{f} \in \mathcal{R}_{\tau}$ has a Wesenheit index in bin $j$, $W_I \in \mathcal{I}_j$, is given as the fraction of training set elements in $\mathcal{R}_{\tau}$ that fall into $\mathcal{I}_j$
\begin{equation}
\mathcal{P}_{\tau, j} = \sum_{i = 1}^{n_{\tau}} \frac{\Theta(W_{I}^{i} \in \mathcal{I}_j)}{n_{\tau}} \, ,
\label{eq:bin_prob_def}
\end{equation}
where the sum runs over all $n_{\tau}$ training set elements in region $\mathcal{R}_{\tau}$. 
Similarity can be optimized by minimizing the Gini criterion $\mathcal{G}_{\tau}$ in region $\mathcal{R}_{\tau}$:
\begin{equation}
\mathcal{G}_{\tau} = \sum_{j = 1}^{n_{\rm bins}} \mathcal{P}_{\tau, j} (1 - \mathcal{P}_{\tau, j}) \, . 
\label{eq:gini_crit}
\end{equation}
Note that $\mathcal{G}_{\tau}$ vanishes if all training set elements in region $\mathcal{R}_{\tau}$ reside in the same Wesenheit bin, and is maximal if they are equally distributed across the Wesenheit bins. In each recursion step we select a binary split in feature space such, that the summed $\mathcal{G}_{\tau}$ over all regions $\mathcal{R}_{\tau}$ is minimized compared with the previous state. The splitting stops if a minimum number of training set objects are located in the respective region, which is a tuning parameter of the model.  

If a new variable star is queried down the tree into region $\mathcal{R}_{\tau}$, the tree returns the bin probabilities $\mathcal{P}_{\tau, j}$, as defined in Eq.~\ref{eq:bin_prob_def}, for each Wesenheit bin $\mathcal{I}_{j}$. For a more in-depth description of the Random Forest algorithm, we refer the interested reader to the literature \citep[e.g.][]{hastie01statisticallearning, Bishop:2006:PRM:1162264}.

\subsection{Evaluation and Metrics}
\label{subsec:evaluation_metrics}
The analysis presented in \S \ref{sec:analysis_results}, applies the Random Forest classifier to each variable star in the considered sample. To ensure that the trained model generalizes well to unseen data, we iterate the splits into training and test set using the $k$-fold cross validation technique. 
The complete dataset is randomly split into $k$ non-overlapping, equal-sized parts, where we use the common choice $k=10$. The model is then trained on all but the first of these partitions and subsequently applied to the held-out partition. The procedure then continues in the same manner with the remaining partitions, where each partition is held out once. In this way we generate an estimate of the conditional density $p(W_I | \mathbf{f})$ for each star in the sample in $k=10$ chunks. We can then evaluate the performance metrics, such as accuracy of the mean and variance.

We measure the prediction quality using the root mean squared error between the true Wesenheit magnitude $W_{\rm true}$ and the predicted Wesenheit magnitude $W_{\rm pred}$ as
\begin{equation}
{\rm RMSE} = \sqrt{MSE} = \sqrt{\frac{1}{N} \sum_{i = 1}^{N} \left(W_{\rm true, i} - W_{\rm pred, i}\right)^2} \, ,
\label{eq:rmse}
\end{equation}
and the mean error (ME) in the prediction as 
\begin{equation}
{\rm ME} = \frac{1}{N} \sum_{i = 1}^{N} \left(W_{\rm true, i} - W_{\rm pred, i}\right) \, .
\label{eq:offset}
\end{equation}
To accurately use variable stars for distance measurements, an important quality requirement is that the variation in the slope of the regression function is small across the several distance anchors \citep[see e.g.][]{2017MNRAS.471.2254Z}. To compare the accuracy of the prediction across samples, we consider the linear regression function
\begin{equation}
W_{\rm true} = b \, W_{\rm pred} + \mu \, ,
\end{equation}
where $\mu$ represents the offset of the linear regression fit and $W_{\rm pred}$ ($W_{\rm true}$) the predicted (true) Wesenheit indices. Since the offset in these relations is connected with the distance modulus between both samples, we compare the similarity between both relations in terms of the relative bias in the slope $b$ between a reference sample $\mathcal{R}$ and a comparison sample $\mathcal{C}$: 
\begin{equation}
b_{\Delta} = \left(b_{\rm \mathcal{R}} - b_{\rm \mathcal{C}}\right) \big/ b_{\rm \mathcal{R}} \, .
\label{eq:rel_bias_b}
\end{equation}

\subsection{Outlier Removal}
\label{subsub:outlier_removal}
Outlier removal is an essential part of accurately using period-luminosity relations of variable stars for distance measurements. A common technique \citep{2017MNRAS.471.2254Z} removes outliers that deviate more than $\alpha \, \sigma$ from the linear $W_{\rm pred}$-$W_{\rm true}$ regression line. In this work we use $\alpha = 2.25$, which is a common choice used in the literature \citep{2017MNRAS.471.2254Z}. Specifically, we iteratively repeat $\alpha\, \sigma$ outlier rejection and linear regression of non-rejected datapoints until we reach convergence.

\section{Analysis and Results}
\label{sec:analysis_results}
In the following section we apply the methodology described in \S \ref{sec:methodology} to the OGLE photometric variable star catalogs of the Magellanic Clouds. We rank the long period variable stars in these samples based on our estimates of their conditional standard deviation (Eq.~\ref{eq:cond_std}). We select variable stars where these values are small, and for which therefore the lightcurve parameters are highly predictive of their Wesenheit. We then evaluate the metrics defined in \S \ref{subsec:evaluation_metrics} on these selections on the selected subsamples in the LMC and the SMC catalogs.

\subsection{Large Magellanic Cloud}
\label{subsec:large_magellanic_cloud}
\begin{figure*}
   \centering  
   \includegraphics[scale=0.55]{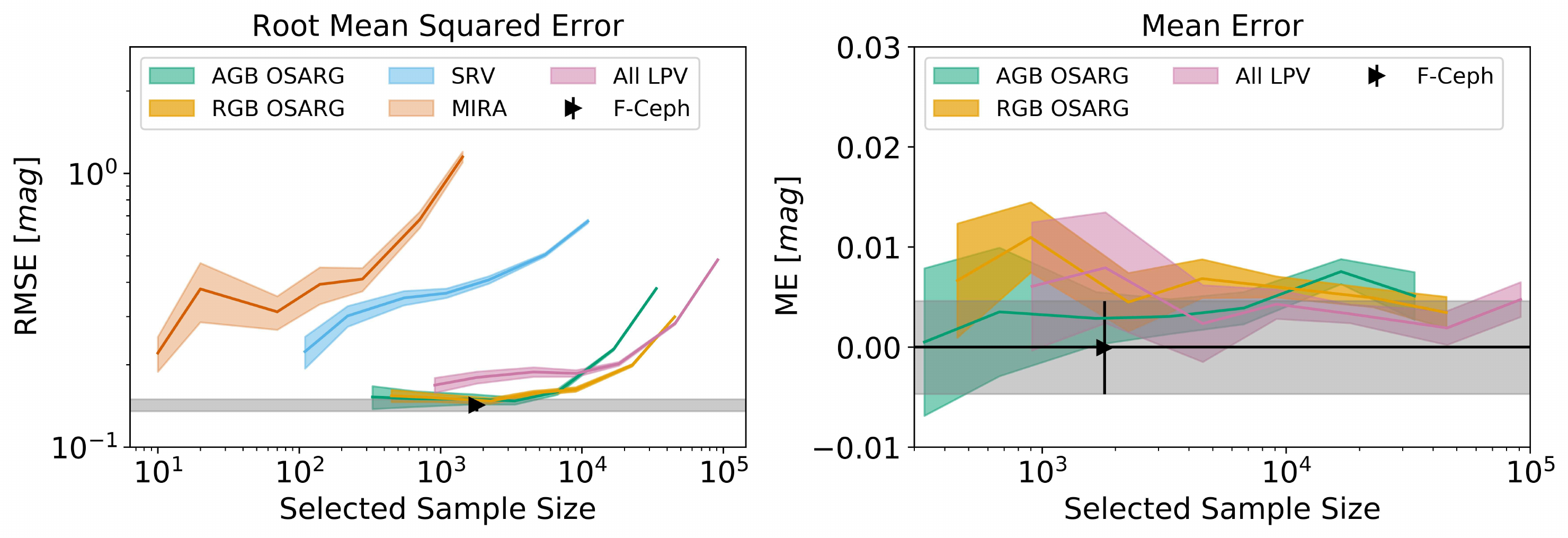}
   \caption{\label{fig:rmse_lmc} Root mean squared error and mean error {\color{black} of the Machine Learning predictions evaluated on different samples of long period variable stars (LPV) in the LMC} as a function of the selected sample size. The horizontal axis shows how many stars are kept in the sample after applying our sample selection methodology (\S \ref{sec:methodology}). In the left panel, the y-axis shows the root mean squared error (Eq. \ref{eq:rmse}) for asymptotic giant branch (AGB) OSARG, red giant branch (RGB) OSARG, SRV and MIRA stars as well as for all LPV. The right panel shows the mean error (Eq. \ref{eq:offset}) for asymptotic giant branch (AGB) OSARG, red giant branch (RGB) OSARG and for all LPV. The horizontal grey contour shows the respective results obtained using fundamental mode Cepheids (F-Ceph), where the black right triangle indicates the Cepheid sample size. The mean and errorbars are obtained by 10-fold cross validation as explained in the text.  } 
\end{figure*}

In this section we investigate how the accuracy of the Random Forest predictions improve when we optimize the sample selection using the methodology described in \S \ref{sec:methodology}. 
To this end, we use the 10-fold cross validation procedure to obtain predictions of the conditional mean (Eq.~\ref{eq:cond_mean}) and conditional standard deviation (Eq.~\ref{eq:cond_std}) for the variable star \markus{samples} in the LMC. For each of the 10 equal-sized folds, we select the variable stars based on the smallest estimated conditional standard deviation and evaluate the performance metrics on this selection. Thus for each selected sample size we obtain 10 cross validated performance estimates. This procedure is separately applied to the full sample of long period variables, and to \sk{four subclasses of variable stars}: AGB/RGB OGLE Small Amplitude Red Giants, Mira variables and semi-regular variables (SRV). The left (right) panel of Fig.~\ref{fig:rmse_lmc} shows the {\color{black} root mean squared RMS error (mean error ME)}\footnote{{\color{black} The root mean squared error (RMSE) and the {\color{black} mean error (ME)} are defined in Eq. \ref{eq:rmse} and Eq. \ref{eq:offset} respectively.}} as a function of the selected sample size, i.e. the number of variable stars that remain in the sample after the selection \sk{by conditional standard deviation $\sigma(W|f)$ as described above}. The solid mean line shows the average performance across the $k=10$ equal sized folds and the error contours the standard deviation across them. \markus{We quote the combined sample size of the 10 folds on the horizontal axis and divide the width of the $1 \sigma$ errorbars by $\sqrt{k}$ to correct for the increased sample size. }

To compare with the performance of a well-known standard candle, we show these metrics, obtained in the same manner, for the fitted Period-Wesenheit (P-W) relation of the LMC Cepheids as a grey band. The black bullet corresponds to the sample size of this Cepheid sample. We note that the AGB and RGB OSARGs show small RMSE values that are consistent with the performance of the Cepheid P-W relation within the statistical errors, while having a factor of three times larger sample size. 

\begin{figure}
   \centering  
   \includegraphics[scale=0.09]{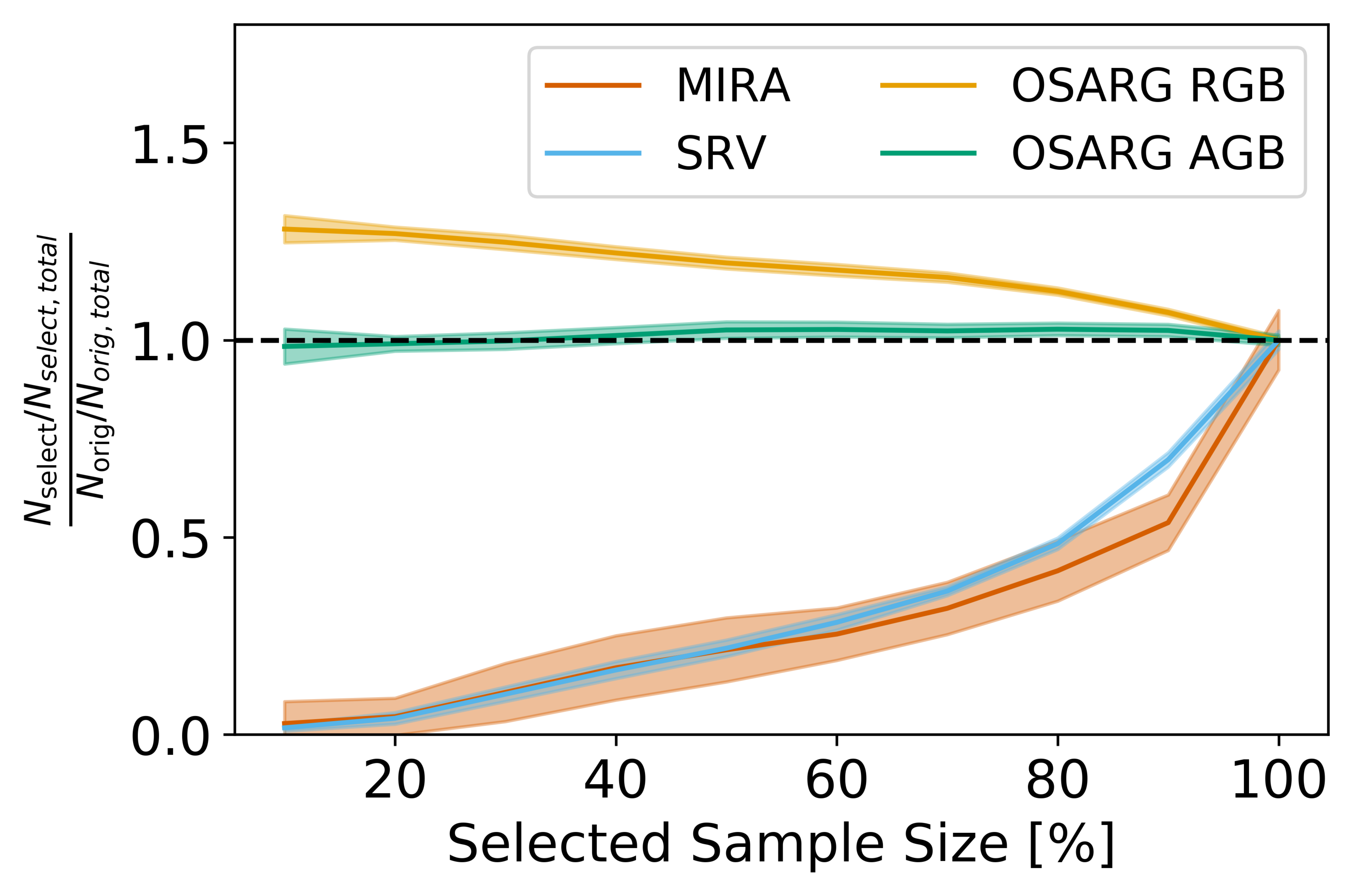}
   \caption{\label{fig:pop_difference} {\color{black} Excess probability over a random selection for the different types of long period variables as a function of the selected sample size in the full LMC LPV sample. The black horizontal dashed line shows the selection probability that would be expected in a completely random selection. The brown, blue, yellow and green lines show the population fractions for the different types of LPV, if the selection is done using our method. }} 
\end{figure}

{\color{black} Fig.~\ref{fig:rmse_lmc} also shows the results of our method when applied to the full LPV sample (`All LPV') with all its subclasses, i.e. Mira, SRV and OSARG variables. The performance in this case is quite similar to AGB OSARG and RGB OSARG. Comparing with the performance of e.g. the SRV population at constant sample size, we see that our methodology robustly identifies subpopulations of LPVs with especially tight conditional pdfs purely based on light curve features, without a prior step of variable star classification. 

We highlight this point in the following analysis: we use the `All LPV' sample, i.e. the sample of all long period variables in the LMC, and investigate how the fraction of different types of LPV variables changes, when we select subsamples using our proposed method. We measure this change as the fractional population difference: 
\begin{equation}
    \Delta_{\rm frac, type} = \frac{N_{\rm select}/N_{\rm select, total}}{N_{\rm orig}/N_{\rm orig, total}} \, .
\end{equation}
Here $N_{\rm select}$ denotes the number of stars of a given type after the selection and $N_{\rm select, total}$ denotes the total number of stars, including all types, after the selection. $N_{\rm orig}$ denotes the number of stars of a given type before the selection, i.e. for the full sample, and $N_{\rm orig, total}$ denotes the total number of stars for the full sample, including all types. This quantity therefore measures how the fraction of a given type of variable stars changes in the sample, when we apply a more restrictive selection.

Fig. \ref{fig:pop_difference} plots $\Delta_{\rm frac, type}$ as a function of the selected sample size. We see that the fraction of MIRA and SRV variable stars decreases significantly, while the fraction of OSARG AGB and OSARG RGB stars remains constant and even increases. We note that the horizontal line shows the expectation of a random selection, where the fraction of LPV types remains constant. The $1\sigma$ errorbars show the deviation between the 10 cross validation folds. }
\begin{figure*}
   \centering  
   \includegraphics[width=0.95\textwidth]{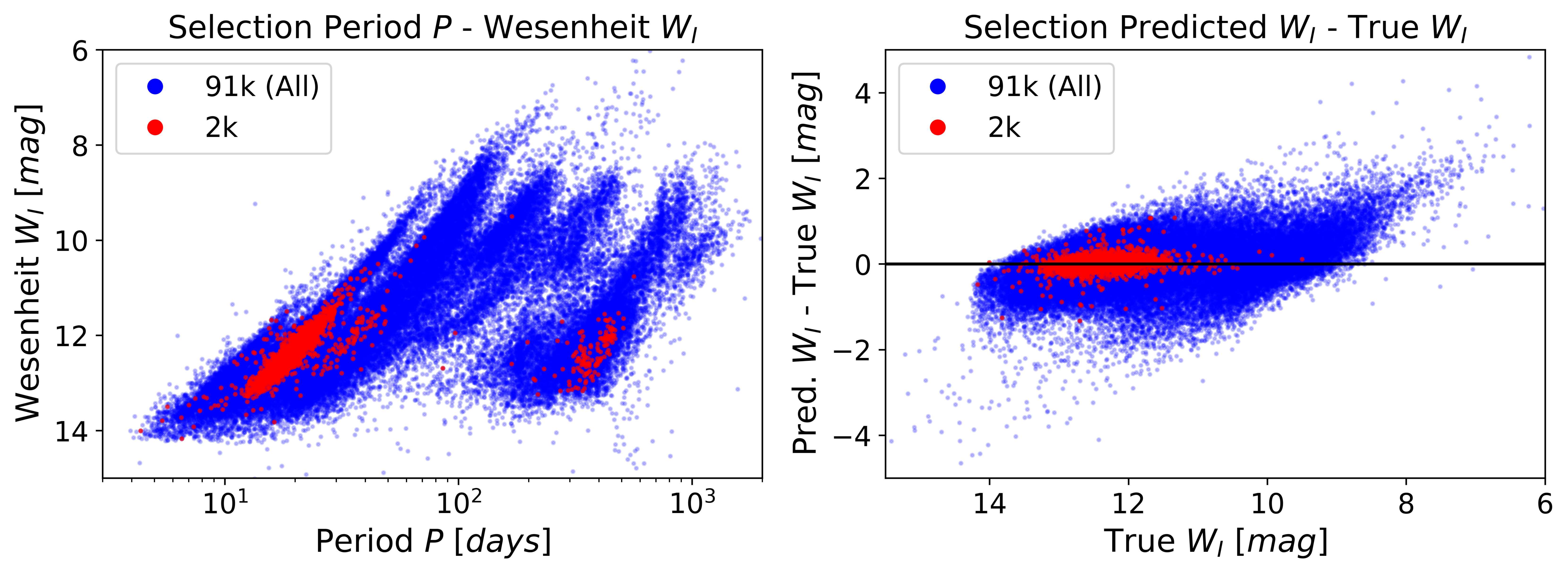}
  \caption{\label{fig:selected_population_lmc} Period-Wesenheit relations and performance of our selected sample of long period variables in the LMC. \textit{Left}: Period $P$ against Wesenheit $W_I$ for long period variables (LPV) in the LMC. Blue points show the full sample, red points show the 2k stars that have the smallest conditional standard deviation. \textit{Right}: Corresponding plot of the true Wesenheit $W_I$ against the difference between the predicted and true Wesenheit $W_I$. The red points again show the 2k predictions for which our model infers the smallest conditional standard deviation.} 
\end{figure*}

Compared with the performance of the OSARG AGB/RGB subsamples, we note that Mira or SRV variables show a substantially larger RMSE, {\color{black} due to a larger luminosity range at fixed period compared with OSARG variables.} We expect however that the inclusion of {\color{black} deep infrared photometry in the K band \citep{2008MNRAS.386..313W, 2017AJ....153..170Y}} will improve the residuals obtained for the Mira sample, as Miras follow a well defined P-W relation in this wavelength range \citep[see e.g.][]{2018arXiv180102711H}. 
Similarly the mean errors obtained for those samples that show a small RMSE, i.e. the AGB OSARG, RGB OSARG and the full LPV sample, are consistent with the results obtained using Cepheids, albeit having a small bias towards underpredicting the true Wesenheit. We note that this result is stable across all considered sample sizes, where the width of the error decreases with the inclusion of more stars to the sample. For large sample sizes of $N > 10.000$ stars, the error on the mean error (ME), shown as $1 \sigma$ contours, obtained on the LPV samples is significantly smaller than the one for Cepheids. {\color{black} At the sample size of the Cepheid sample, these errorbars are comparable between Cepheids and LPVs.}

The left panel of Fig.~\ref{fig:selected_population_lmc} color codes the selected long period variable stars in the P-W relation. We see that the best 2000 long period variables\footnote{For this sample size, the `All LPV' sample in Fig. \ref{fig:rmse_lmc}, is roughly of comparable size to the Cepheid sample and has similar prediction accuracy.}, cluster around the faint part of the respective Wood sequences, in a tight correlation between log period and Wesenheit. This result again highlights that the algorithm is able to effectively select samples of variable stars solely based on the provided input features without the need of a prior classification step. In addition we note that the selected sample is on average significantly brighter than the Cepheid sample. While the Cepheid sample has Wesenheit of $11 \lesssim W_{I} \lesssim 16$, the selected LPV sample span a range of $11 \lesssim W_{I} \lesssim 13$ (see Fig.~\ref{fig:selected_population_lmc}). \markus{Their brightness makes these LPV stars attractive as potential standard candles for cosmological distance measurements.} We note that imposing a less stringent sample selection on $\sigma(W|\mathbf{f})$ will allow more and even brighter objects into the sample. We discuss this in more detail in \S \ref{subsec:selected_population}. The right panel of Fig.~\ref{fig:selected_population_lmc} plots the true Wesenheit against the difference between predicted and true Wesenheit for both the full sample and our selection. We see that the performance of the selected, or optimized, sample is much better compared with the full sample. {\color{black} We also note that the multimodality in Wesenheit magnitude at a given period as shown in Fig.~\ref{fig:selected_population_lmc}, is removed and the Wesenheit distribution as estimated by the Random Forest algorithm given the full feature set is now more peaked and unimodal. }

\subsection{Small Magellanic Cloud}
\label{subsec:cross_train_smc}
In the previous analysis we trained our model exclusively on LMC data using the 10-fold cross validation procedure. In order to test the robustness of our model on other datasets, we now train our model on LMC data and apply it to the corresponding SMC datasets. Since we demonstrated in the previous section that the OSARG stars are the best-performing subpopulation of LPV in the LMC, we concentrate in this section on this subtype for simplicity. In the following section we will train separately on the AGB and RGB subpopulation.

We use the Random Forest models trained on the 10 cross validation folds of AGB/RGB OSARG variables in the LMC and query the corresponding SMC datasets. \markus{We therefore obtain 10 sets of predictions for the SMC AGB and  SMC RGB population respectively, each corresponding to a cross validation fold in the LMC.} The variance between these 10 predictions therefore quantifies the variance in the training of the Random Forest. Since the Random Forest is trained on the LMC data, the predicted Wesenheit magnitudes obtained on SMC data will be biased low due to the distance modulus between LMC and SMC. We can measure this offset by fitting linear regression functions to the selected relations between predicted $W_{\rm pred}$ and true Wesenheit $W_{\rm true}$ for both the LMC and SMC samples. To mimic a typical distance measurement procedure using variable stars, we apply the outlier removal algorithm described in \S \ref{subsub:outlier_removal} to each set of SMC predictions and to each of the 10 cross-validated LMC prediction sets. In analogy to the previous section, we compare the performance of our methodology with the performance of the P-W fits of the fundamental mode Cepheid sample in the LMC and SMC. To make a comparison easier we do not consider the P-W relation directly but instead consider the predicted Wesenheit indices from the fitted P-W relation. 

\begin{figure*}
   \centering  
   \includegraphics[scale=0.065]{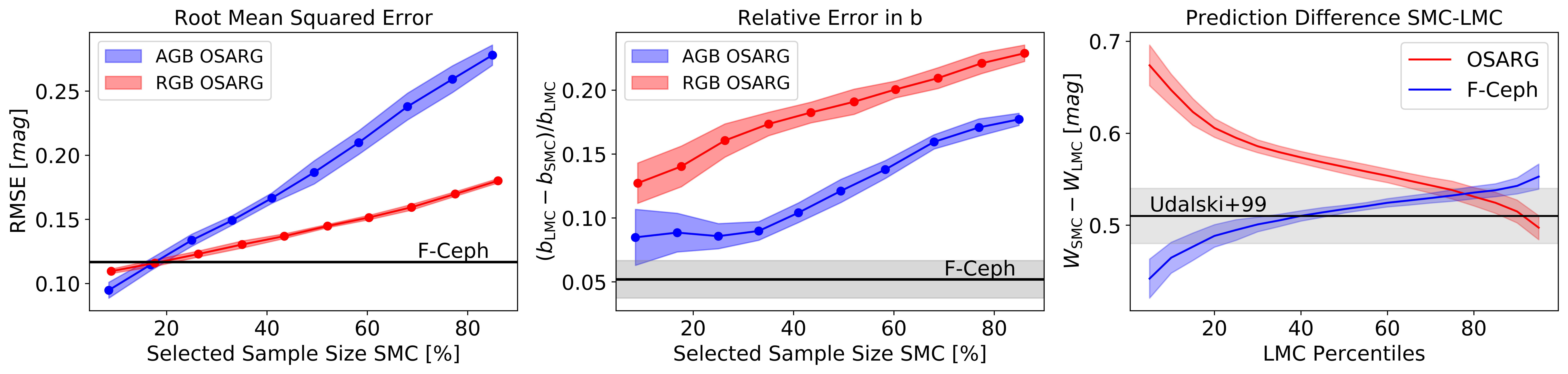}
   \caption{\label{fig:smc_metrics} Metrics to quantify the generalization performance of the Random Forest predictions. The model is trained on LMC data and applied to the respective SMC data. Then a linear regression is fitted between the predicted and true Wesenheit magnitudes. The errorbars shown in these plots are the $\pm 1 \sigma$ standard deviations from the 10 cross validation folds as described in \S \ref{subsec:cross_train_smc}. The quoted fractional sample sizes correspond to the `SMC AGB' and `SMC RGB' columns of Tab.~\ref{tab:outlier_removal}. \textit{Left}: Root mean squared error for the linear regression obtained on the SMC predictions as a function of the remaining fraction of SMC data (\S \ref{subsec:cross_train_smc}). We show the result for the AGB OSARG (blue), the RGB OSARG (red) and the corresponding result from the fundamental mode Cepheids (F-Ceph). \textit{Middle:} Relative difference in the slope parameter of the linear regression functions fitted on the LMC and SMC samples, as a function of the remaining fraction of LMC and SMC data respectively. For the F-Ceph result, we plot the absolute value of this metric for easier comparison. \textit{Right}: \markus{Offset} between the linear regression functions fitted on the LMC and SMC samples. The result is shown for the OSARG AGB stars that correspond to an original selection of 20\% (see Tab.~\ref{tab:outlier_removal}) and the sample of fundamental mode Cepheids. The horizontal axis grid is given by the LMC percentiles of predicted Wesenheit magnitude for the OSARG AGB and fundamental mode Cepheid sample respectively. The horizontal grey area corresponds to the $\pm 1 \sigma$ error contour of the measured difference of distance moduli between LMC and SMC using OGLE fundamental mode Cepheids by \citet{1999AcA....49..201U}.    } 
\end{figure*}

In analogy to the {\color{black} Machine Learning (ML) approach}, we fit the P-W relation on the LMC and apply the model to the corresponding SMC data. Note that the outlier rejection and 10-fold cross validation\footnote{\markus{Our fundamental mode Cepheid sample in the LMC (SMC) contains 1806 (2603) stars from which $89 \pm 3$ \% ($89 \pm 0$\%) remain after the outlier rejection method is applied.}} are analogously applied to the Cepheid sample to obtain error contours in the quoted metrics. The outlier rejection algorithm will remove a certain fraction of the data after the cut on the conditional standard deviation is performed. We will refer to the sample size that includes this cut but not the outlier rejection step as the `original selection' in the following. The exact number of stars being culled depends on the subsample and, to a much lesser degree, on the cross validation fold. We report these numbers in Tab.~\ref{tab:outlier_removal}. Since we are mostly interested in the properties of the SMC sample in this section, we will use the corresponding SMC fractional samples size (marked in boldface) as a reference in Fig.~\ref{fig:smc_metrics}. The corresponding LMC sample sizes can be read-off from Tab.~\ref{tab:outlier_removal}. 
\begin{table}{
\begin{tabular}{c|c|c|c|c}
 Orig &  LMC AGB & SMC AGB  &  LMC RGB   & SMC RGB \\\hline\hline
   $10 \, \%$  &  $9.3 \pm 0.1  \, \%$   & $\mathbf{8.5} \pm 0.2  \, \%$    & $9.0 \pm 0.2  \, \%$ &  $\mathbf{9.0} \pm 0.1  \, \%$   \\ 
   $20 \, \%$  &  $18.6 \pm 0.3 \, \%$  & $\mathbf{16.8} \pm 0.4 \, \%$   & $18.1 \pm 0.3 \, \%$ & $\mathbf{17.7} \pm 0.1 \, \%$   \\
   $30 \, \%$  &  $27.4 \pm 0.3 \, \%$  & $\mathbf{25.0} \pm 0.3 \, \%$   & $27.0 \pm 0.3 \, \%$ & $\mathbf{26.3} \pm 0.2 \, \%$   \\ 
   $40 \, \%$  &  $35.6 \pm 0.6 \, \%$  & $\mathbf{33.1} \pm 0.6 \, \%$   & $35.8 \pm 0.4 \, \%$ & $\mathbf{35.0} \pm 0.2 \, \%$   \\
   $50 \, \%$  &  $43.5 \pm 0.6 \, \%$  & $\mathbf{41.0} \pm 0.5 \, \%$   & $44.4 \pm 0.5 \, \%$ & $\mathbf{43.5} \pm 0.2 \, \%$   \\
   $60 \, \%$  &  $51.0 \pm 0.7 \, \%$  & $\mathbf{49.4} \pm 0.7 \, \%$   & $52.8 \pm 0.7 \, \%$ & $\mathbf{52.1} \pm 0.2 \, \%$   \\
   $70 \, \%$  &  $58.8 \pm 0.9 \, \%$  & $\mathbf{58.3} \pm 0.9 \, \%$   & $60.9 \pm 0.9 \, \%$ & $\mathbf{60.3} \pm 0.2 \, \%$  \\
   $80 \, \%$  &  $66.4 \pm 1.5 \, \%$  & $\mathbf{68.0} \pm 1.1 \, \%$   & $68.4 \pm 0.9 \, \%$ & $\mathbf{68.8} \pm 0.3 \, \%$ \\
   $90 \, \%$  &  $73.1 \pm 1.4 \, \%$  & $\mathbf{77.0} \pm 1.1 \, \%$   & $75.7 \pm 1.1 \, \%$ & $\mathbf{77.5} \pm 0.3 \, \%$  \\
   $100 \, \%$ & $78.9 \pm 1.6 \, \%$   & $\mathbf{84.9} \pm 1.0 \, \%$   & $82.6 \pm 1.3 \, \%$ & $\mathbf{86.0} \pm 0.3 \, \%$  \\
\end{tabular}
\caption[Outlier removal]{ Reduction in sample size due to outlier removal in percent. The first column shows the fractional sample size after imposing the cut on the conditional standard deviation, `original selection' in the following. The other columns report the corresponding fractional sample sizes after the outlier removal for the different subsamples. The SMC results are shown in boldface and will be used as reference in this analysis. The $1\sigma$ errors quantify the variation across the 10 cross validation folds. }
\label{tab:outlier_removal}}
\end{table}

{\color{black} We quantify the performance of our methodology relative to the Cepheid P-W fit in Fig.~\ref{fig:smc_metrics}.
In the left panel we show the RMSE (Eq.~\ref{eq:rmse}) as a function of the culled fraction of SMC variable stars for the RGB and the AGB population and compare these values with the corresponding result for the P-W fit of fundamental mode Cepheids (F-Ceph) in the SMC. 
We note that both the RGB OSARG as well as the AGB OSARG population show a rapid decline in RMSE when more data is being culled. While the RGB OSARG sample has a smaller RMSE than the AGB OSARG sample for an original selection of 100\%, both samples show quite similar results to each other and to F-Cepheids for an original selection of 20\%. 
The middle panel of Fig.~\ref{fig:smc_metrics} shows the relative bias in the slope of the linear regression fits (Eq.~\ref{eq:rel_bias_b}) as a function of the culled fraction of the respective sample. We see that the OSARG AGB population performs much better in terms of this metric as compared with the RGB sample. For an original selection of 20\%, the performance of the AGB subsample is comparable with the Cepheid sample reference within the $1 \sigma$ errorbars.

The right panel concentrates on the AGB OSARG and the Cepheid sample. We show the offsets between the linear regression functions estimated on the SMC and the LMC samples $W_{\delta} = W_{\rm pred, SMC} - W_{\rm pred, LMC}$, evaluated at the percentiles of the predicted Wesenheit distribution of the respective LMC OSARG AGB and LMC fundamental mode Cepheid sample. This allows us to compare both samples, that cover a different range in Wesenheit index, on the same scale. We note that the slightly larger relative error in the regression slope $b_{\Delta}$ obtained on OSARG AGB stars propagates into a larger variation in $W_{\delta}$, compared with the Cepheid reference. However both results are comparable and overlap at the faint end, i.e. at the 80th percentile. We compare these results with measurements of the LMC-SMC distance modulus by \citet{1999AcA....49..201U} that used a comparable sample of OGLE Cepheids. The corresponding horizontal grey $1 \sigma$ error contours are consistent with our Cepheid results and also with the results obtained using AGB OSARGs at the faint end. Comparing with Fig.~\ref{fig:rmse_lmc}, we note that the error induced by biases in the regression slope ($\approx 0.1 \, {\rm mag}$) is an order of magnitude larger than the mean error in the LMC distance anchor ($\approx 0.01 \, {\rm mag}$). Controlling the bias in the regression slope is therefore the most important challenge for obtaining more accurate distance measurements.

\begin{figure*}
   \centering  
   \includegraphics[width=0.95\textwidth]{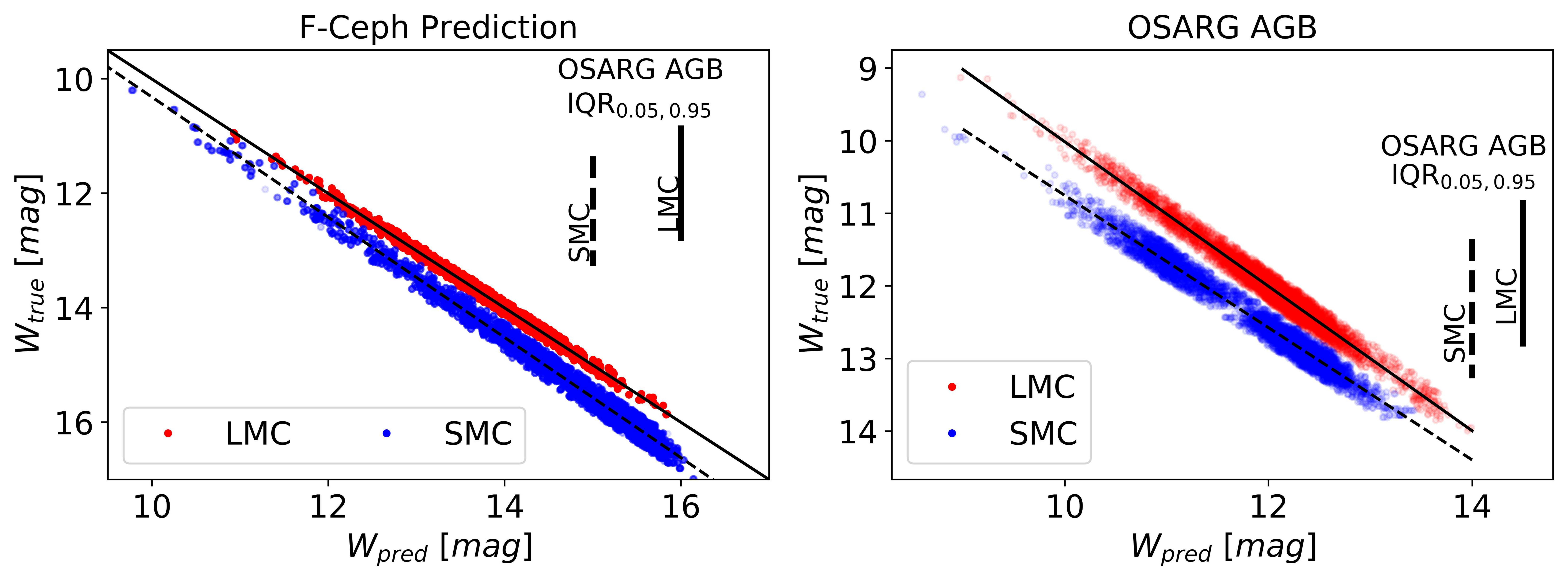}
   \caption{\label{fig:scatter_plot_lpv} Performance when training on LMC data and applying the model to both LMC and SMC data (\S \ref{subsec:cross_train_smc}) for an original selection of 20\%, \markus{i.e. when beginning the analysis with 20\% of the sample based on the cut on the conditional standard deviation, before further outlier rejection.} \textit{Left}: Wesenheit prediction for fundamental mode Cepheids (F-Ceph) in the LMC (red) and SMC (blue) datasets. The dashed (solid) vertical lines show the inter-quantile range of true Wesenheit values covered by the OSARG AGB sample for the SMC (LMC) datasets. \textit{Right}: Wesenheit predictions for an original selection of 20\% OSARG AGB stars for the LMC (red) and SMC (blue). } 
\end{figure*}
Fig.~\ref{fig:scatter_plot_lpv} shows the resulting $W_{\rm pred}$-$W_{\rm true}$ relations for the Cepheid population (left panel) and the population of OSARG AGB stars (right panel) selected using the best 20\% elements from the original selection based on their conditional standard deviation\footnote{This corresponds to the maximum sample size in the left and middle panel of Fig.~\ref{fig:smc_metrics}, where the RMSE and bias in the slope $b_{\Delta}$ for the OSARG AGB sample is comparable with the Cepheid result.}. For simplicity, in this plot, we merge the datasets in the 10 cross validation folds before applying our methodology.
Comparing both panels of Fig.~\ref{fig:scatter_plot_lpv}, we note that the OSARG AGB sample is at the bright end of the fundamental mode Cepheid (F-Ceph) sample as shown by the vertical lines that indicate the inter-quantile range $\rm{IQR}_{0.05, 0.95}$ between the 0.05 and the 0.95 quantile of the selected true OSARG AGB Wesenheit distribution. We see that the regression lines between the LMC and the SMC are slightly biased for both the Cepheid and the OSARG AGB prediction. We also note that this bias is complementary for both samples, i.e. the offset between the respective SMC and LMC linear regression lines is larger at the bright end for the OSARG AGB subsample and decreases towards the faint end, whereas the contrary is the case for the Cepheid sample. This is also highlighted in the right panel of Fig. \ref{fig:smc_metrics} that shows the distance between the LMC and SMC linear regression lines as a function of the predicted LMC Wesenheit percentiles. High (low) percentiles indicate the faint (bright) end of the OSARG AGB and F-Ceph predicted LMC Wesenheit distributions. This complementarity of systematic biases suggests that combining distance measurements using both samples might compensate for their respective systematic errors. This will however require that the respective samples are sufficiently complete to avoid introducing additional sample selection biases. We leave a detailed study of this for future work. }

\subsection{Selected Population}
\label{subsec:selected_population}
\begin{figure*}
   \centering  
   \includegraphics[width=0.95\textwidth]{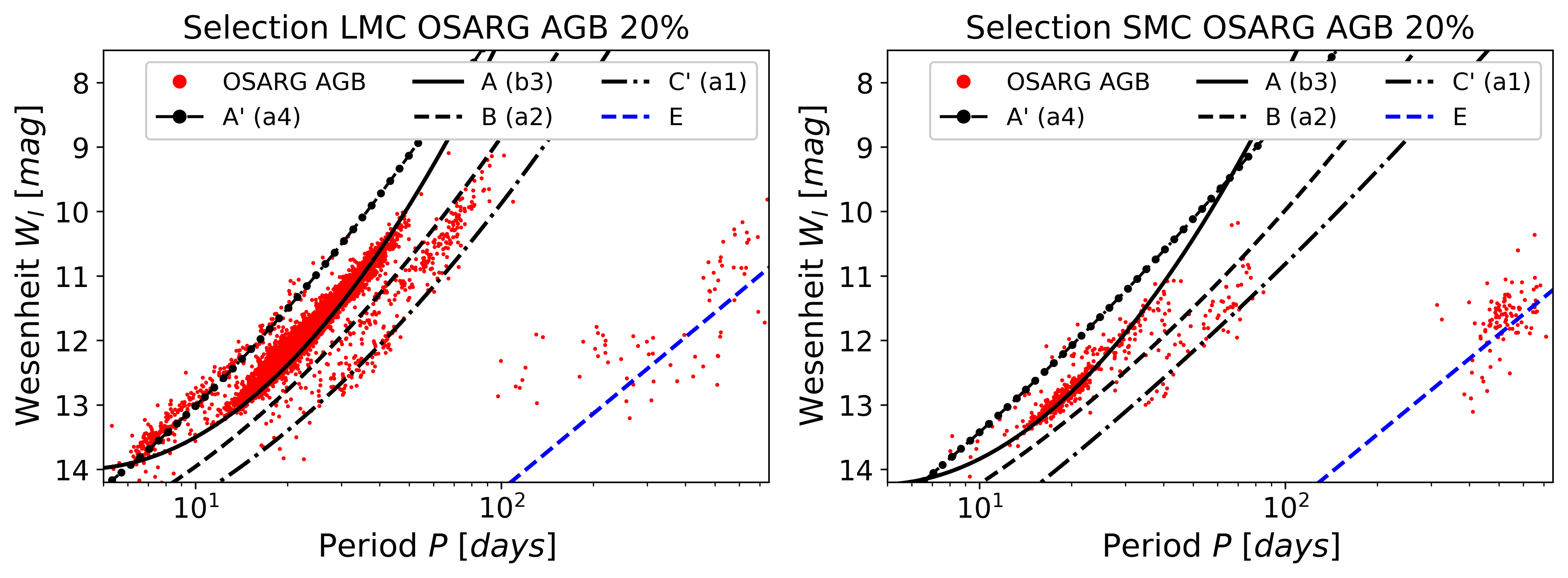}
   \caption{\label{fig:pop_overview} P-W relation of the selected AGB OSARGs in the LMC and SMC. \textit{Left}: P-W relation of the 20\% best OSARG variables selected in the LMC (red points). We overplot the Wood sequences obtained from \citet[][table~1]{2007AcA....57..201S} that best fit our selection. The legend quotes the sequences according to the two conventions in \citet{2017ApJ...847..139T} and, in parenthesis, \citet{2007AcA....57..201S}. \textit{Right}: Corresponding plot for the SMC selection and Wood sequences extracted from \citet[][Tab. 2]{2007AcA....57..201S}. We note that no outlier rejection has been applied.} 
\end{figure*}
The previous sections demonstrated that our methodology has potential to incorporate LPVs into distance measurements. In \S \ref{sec:data} we discussed that LPVs, including OSARG variables, follow tight P-W relations and exhibit characteristic period ratios that will help the machine learning algorithm to distinguish between them. The incorporation of oscillation amplitudes will also help to improve the performance of both the selection as well as the final prediction, as they relate to the growth rate of oscillation modes and the luminosity of the star. 

In this section we discuss which stars are preferentially selected by our methodology, on the example of the OSARG AGB variables. The left (right) panel of Fig.~\ref{fig:pop_overview} shows the P-W relation of the 20\% best LMC (SMC) OSARG variables selected by our methodology. We note that no outlier rejection is applied here, as we want to show the selection based purely on our methodology. The plot shows that the selected stars cluster around tight `Wood-sequences' \citep{1999IAUS..191..151W, 2000PASA...17...18W, 2004MNRAS.347..720I, 2015MNRAS.448.3829W, 2017ApJ...847..139T}, that correspond to different overtones of stellar oscillation. Overplotting the analytical sequences obtained on the LMC OGLE-III dataset by \citet{2007AcA....57..201S}, we can attribute our selected sequences, starting from the lower period end, to the $A'$, $A$, $B$/$C'$ and $D_c$/$E$ relations. \citet{2017ApJ...847..139T} analyzed these sequences using a pulsation and stellar population synthesis model tuned to resemble the population of red giants in the LMC. Following their discussion the selected sequences $A'$ and $A$ {\color{black} are attributed} to the third (O3) and second (O2) overtone, and the splitted sequences $B$/$C'$ to the first overtone (O1). The {\color{black} long period} population can be roughly associated with the sequences $D_c$ and $E$ in \citet{2007AcA....57..201S}, which could potentially be caused by stellar pulsation \citep{2004ApJ...604..800W, 2015MNRAS.452.3863S, 2017ApJ...847..139T} or binarity\footnote{A low mass companion `drags' a dust cloud ejected from the central red giant star, which disturbs its lightcurve.} \citep{2004ApJ...604..800W, 2007ApJ...660.1486S}. The right panel of Fig.~\ref{fig:pop_overview} shows the corresponding selection for the SMC OSARG AGB sample. {\color{black} We note that the selections between both Magellanic clouds populate the same Wood sequences. Accordingly we can assume that our selection generalizes well to the SMC, as stars with very similar oscillation patterns are selected. This explains the good performance of our model on the SMC data as demonstrated in the previous section.}

\section{Summary and Conclusions}
\label{sec:conclus}
This paper introduced a novel methodology to select variable stars based on the width of the posterior distribution of their Wesenheit magnitude given their lightcurve parameters. Our selection procedure uses the Random Forest algorithm to estimate this conditional predictive distribution $p(W_I | \mathbf{f})$ of the Wesenheit $W_I$ given a set of features $\mathbf{f}$ extracted from the lightcurve of the variable star. Our feature set $\mathbf{f}$ consists of the first three periods and oscillation amplitudes from the fourier lightcurve fit. We then select variable stars that have a small standard deviation in the conditional predictive distribution. 

This selection procedure constructs a sample of variable stars, that show a very tight correlation between the extracted lightcurve features and their Wesenheit magnitude, which is an important requirement to use them as cosmological standard candles. We demonstrate the effectiveness of this methodology using samples of variable stars in the Large Magellanic Cloud (LMC) and the Small Magellanic Cloud (SMC), observed in the photometric bands $I$ and $V$ by the OGLE collaboration. We show that our method is able to select a subsample of variable stars within the LMC for which highly accurate predictions of the Wesenheit magnitude, as quantified by the root mean squared error (RMSE), can be derived. For the sample of OGLE Small Amplitude Red Giant (OSARG) stars, we show that the RMSE of these predictions \markus{are} comparable to the results obtained on fundamental mode Cepheids in the LMC. However the sample of OSARGs with comparable RMS error is larger by a factor of 3-4 and brighter by $\approx 2$ Wesenheit magnitudes on average. 
Accordingly, we can select a sample of variable stars with comparable systematics to Cepheids, that is both more numerous and brighter. This provides exciting prospects to extend the distance ladder to extragalactic galaxies by utilizing these variable stars to improve the calibration of local supernovae samples in distance anchors such as the LMC or M31. 

To demonstrate the generalization performance on unseen data from a different galaxy, we train our model on the OSARG sample in the LMC and apply the trained model to the corresponding OSARG sample in the SMC. Using our selection methodology, we obtain RMSE values consistent with the corresponding results obtained by fitting the Period-Wesenheit relation (P-W) on fundamental mode Cepheids. Furthermore we investigate how the bias in the slope of the resulting linear regression between the predicted and the true Wesenheit differs between the samples. This quantity is of specific interest as it contributes to the systematic error budget in the distance calibration of local supernovae. We find that for the AGB subsample we can obtain biases that are comparable with the results obtained using fundamental mode Cepheids. The distance modulus between the LMC and the SMC regressions obtained using fundamental mode Cepheids and OSARG AGB variables is consistent at the faint end of the covered range of Wesenheit magnitudes, despite the slightly larger bias in the regression slope as measured on the selected OSARG AGB stars. Notably, \markus{the sign of} this systematic is complementary between the OSARG AGB and fundamental mode Cepheid sample. While at the bright end the regression lines are farther apart for the selected OSARG AGB sample, the contrary is true for the fundamental mode Cepheids. 

This result indicates that there is potential to combine distance measurements obtained using multiple types of variable stars, like Cepheids and OSARG variables, to better control systematic errors. The inclusion of additional photometric bands and lightcurve features will likely improve these results but also allow the incorporation of other types of variable stars like Mira variables that show tight PL relations when observed with deep near infrared photometry \citep{2018arXiv180102711H}. The inclusion of Gaia parallaxes can naturally substitute the apparent Wesenheit magnitudes as regression targets, which will improve the calibration of distances to the LMC and SMC. \markus{While the presented methodology does not require prior variable star classification, it will still be useful to generate additional features that help the methodology to better separate the different subclasses. For example we found that AGB OSARGs in general yielded better results in the considered performance metrics, as compared with the other types of LPV.} 

{\color{black} It has to be noted however, that using OSARG variables in different anchor samples like M31 will require better photometry than Cepheid observations. The small amplitude of OSARG variability sets stringent requirements on the quality of the photometry. The median amplitudes for the selected LMC population in Fig. \ref{fig:pop_overview} range from $0.006 - 0.01 \, [{\rm mag}]$. Thus, the photometric error of the observations has to be of that order to avoid sample selection biases. The observation of long period variables will also require long time series to obtain accurate period estimates. While OSARG variables can have very long oscillation periods of $10^3$ days, the bulk of the stars selected by our methodology have periods $< 100$ days. As a result, the majority of variable stars that would be interesting for distance measurements have period lengths that are comparable to the long period tail of Cepheid samples observed in possible anchor galaxies like M31 \citep[e.g.][]{2018AJ....156..130K}. While OSARGs have a more complicated oscillation pattern than these Cepheids, we still expect that the main observational challenge will be their detection at larger distances. Assuming that these observational requirements can be met, we found that the prediction accuracy is mostly sensitive to accurate measurements of the periods and relatively robust against errors in the amplitudes. This can be explained by the tightly spaced period ratio structure of OSARG variables shown in Fig. \ref{fig:petersen_diagram} and the comparatively broad correlation between amplitude and Wesenheit (see Fig. \ref{fig:features}). We refer for a more discussion to the appendix. In future work we will further investigate which type of LPV stars can be used for distance calibration in fainter samples. For the calibration of local Cepheid distances, OSARG variables already appear as a viable option. }

In a practical application it will also be important to optimize the sample not only with respect to the width of the conditional predictive distribution, but also such that differences in the linear regression slope between the different anchor samples are reduced. This procedure will naturally include a more advanced outlier rejection methodology, that was not optimized in this work. In this context we want to \markus{reiterate} that the methodology presented in this paper does not require a prior classification of the sample into different types of variable stars. Samples of variable stars that are `good standard candles' will be automatically selected in high-dimensional feature space based on their small RMS error and their robust regression functions across different anchor samples. This not only reduces the need to obtain large, accurately labeled training sets to use in variable star classification pipelines, but also avoids biases due to misclassification errors. In future work we will apply our methodology to additional anchor samples and optimize our selection criteria towards the specific science goal of calibrating local supernovae samples.

We conclude that the usage of other types of variable stars \markus{besides Cepheids} as `standard candles' has great potential to improve distance measurements and the calibration of local supernovae samples, which will ultimately lead to a better understanding of sources of systematic error in $H_0$ measurements.

\section*{Acknowledgements}

MMR thanks Ben Hoyle and Kerstin Paech for useful discussions and comments on the draft. \\MMR is supported by DOE grant DESC0011114.  HT and RM are partially supported by NSF grant IIS-1563887. SK is partially supported by NSF grant  AST-1813881.




\bibliographystyle{mnras}
\bibliography{bibliography.bib} 




\appendix
{\color{black}
\section{Feature Importance and Robustness}
The selection methodology presented in this paper is quite general and able to identify samples of variable stars with better prediction accuracy across a wide range of variable star types as shown in Fig. \ref{fig:rmse_lmc}. However we found that Ogle Small Amplitude Red Giant variables show especially tight Period-Wesenheit relations. To understand the sensitivity of these predictions on the input feature set, this appendix performs a feature importance study and tests the robustness of our methodology against feature noise.
\subsection{Feature Importance}
\label{app:feat_imp}
We use the Gini criterion \citep{breiman1984classification} (Eq. \ref{eq:gini_crit}) as implemented in the scikit-learn package \citep{scikit-learn}, as a measure of feature importance. Considering a single tree, we associate a node with a split on a certain variable. The split is selected such that the Gini criterion is decreased. Furthermore, we can attribute a weight to each node defined by the fraction of samples that reach this node. Using these weights, we can average the decrease in the Gini criterion across all trees in the Random Forest to obtain a measure of how important a certain feature is.

Focusing on the LMC OSARG AGB dataset for simplicity, we run the 10 fold cross validation procedure described in \S \ref{subsec:large_magellanic_cloud} to obtain 10 estimates of feature importance as shown in the left panel of Fig.~\ref{fig:feat_imp}. The errobars on the histograms denote the $3 \sigma$ errors across the 10 folds. 

We see that the sets of periods are significantly more important than the sets of amplitudes. The most important features are the first $P_1$ and third $P_3$ period and the secondary amplitude $A_2$. The least important one is the third amplitude $A_3$. We note however that especially the amplitudes are correlated and we therefore expect a level of redundancy in the information contained in the amplitude features. Nonetheless, this result indicates that amplitude information is important for the prediction accuracy, even for the single population of LMC OSARG AGB stars, for which small oscillation amplitudes is a common feature. The much larger importance of periods for the prediction accuracy is to be expected, based on the fine grained period ratios apparent in the Petersen Diagram (see Fig.~\ref{fig:petersen_diagram}) and the strong Period-Wesenheit correlation in the Wood sequences (see \S \ref{subsec:selected_population}).
\begin{figure*}
   \centering  
   \includegraphics[width=0.95\textwidth]{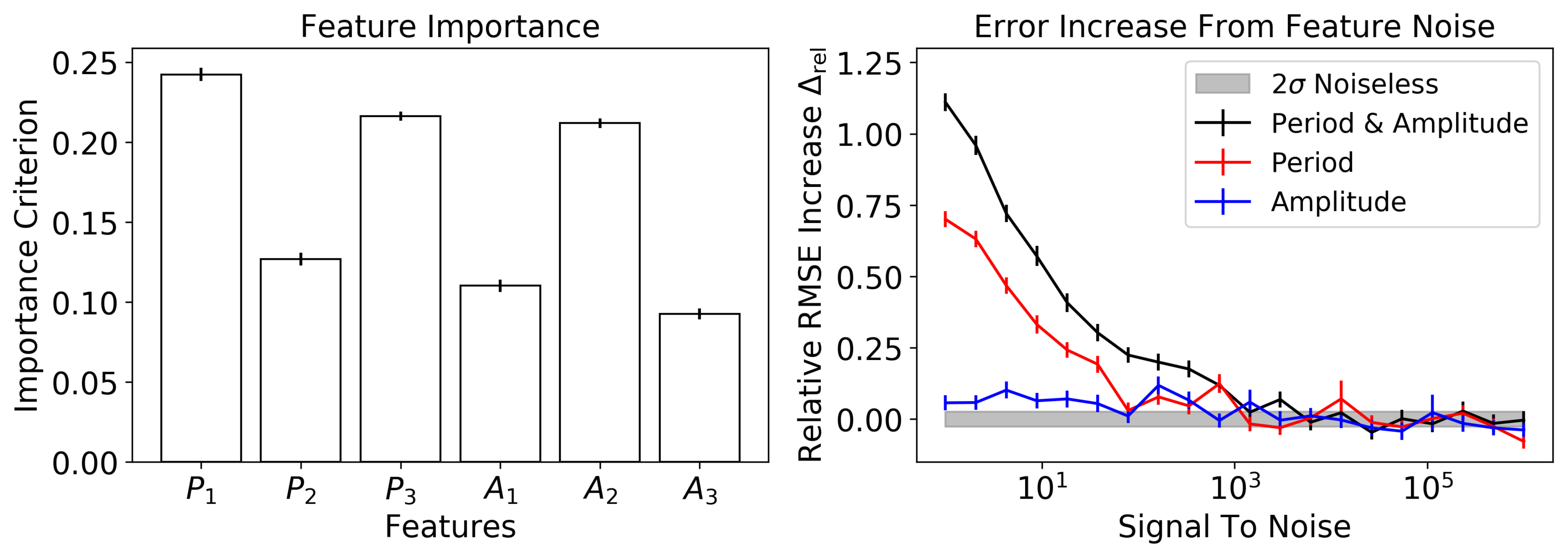}
   \caption{\label{fig:feat_imp} Feature importance and robustness to input noise. \textit{Left:} Feature importance for the full LMC OSARG AGB sample for the period $P_{1-3}$ and amplitude $A_{1-3}$ features used in the analysis. We show the $3 \sigma$ errors on the histograms to quantify the statistical noise.  \textit{Right:} Signal to noise imposed on the respective feature combination against the relative RMSE increase over the noiseless performance for a selected sample of 3038 LMC OSARG AGB stars. The grey horizontal contour and the error bars show the $2\sigma$ error to be expected from the statistical variance. We show the case of imposing noise on the first three periods (red), Amplitudes (blue) and both features (black).  } 
\end{figure*}
\subsection{Feature Noise}
\label{app:feature_noise}
We test the robustness of our Machine Learning methodology with respect to feature noise by artificially imposing Gaussian random noise on the features and subsequently testing the prediction accuracy. For this we perform an analysis similar to \S \ref{subsec:large_magellanic_cloud}. We split the full dataset of LMC OSARG AGB stars randomly into 2 disjoint, similarly sized parts: the training set with 50000 stars and the test set with 41374 stars. We then add Gaussian noise to the set of periods, the set of amplitudes and both sets of features respectively, train our model on the training set and estimate conditional density functions for all stars in the test set. We do this for the noiseless dataset, as well as the three aforementioned combinations of noisy data and perform the subsequent analysis separtely on each of them. We select the 7\% (3038) stars\footnote{This number amounts approximately to the sample size, where the LMC AGB OSARG prediction accuracy begins to significantly deviate from the Cepheid baseline as shown in Fig.~\ref{fig:rmse_lmc}.} from the test set predictions that have the smallest predicted conditional standard deviation. For this selected subset, we estimate the relative increase in the root mean squared error $\Delta_{\rm rel}$ over the noiseless case defined as
\begin{equation}
    \Delta_{\rm rel} = \frac{{\rm RMSE}_{\rm noisy} - {\rm RMSE}_{\rm noiseless}}{\rm RMSE_{\rm noiseless}} \, ,
    \label{eq:rel_inc_rmse}
\end{equation}
where ${\rm RMSE}_{\rm noisy}$ and ${\rm RMSE}_{\rm noiseless}$ denote the root mean squared error (RMSE) defined in Eq. \ref{eq:rmse} for the noisy and noise free datasets. The standard deviation of the Gaussian noise that is imposed on the features is given as $\sigma = \frac{f}{\rm S/N}$, where $\rm S/N$ denotes the signal to noise ratio.
We note that this validation procedure is a slightly simplified version of the 10 fold cross validation approach, as it considers only a single train/test set split. This simplification is justified, as the LMC OSARG AGB sample is quite large even after selecting the best variable stars. For each prediction we obtain errors by propagating the accuracy in the mean squared error ${\rm MSE}$ through Eq. \ref{eq:rmse} and Eq. \ref{eq:rel_inc_rmse}.

The right panel of Fig.~\ref{fig:feat_imp} plots the signal to noise ratio against the relative increase in the RMSE for the three scenarios, i.e. degrading the three amplitudes (blue), the three periods (red) and both the periods and amplitudes (black). The errorbars show the $2 \sigma$ errors. The grey horizontal band shows the $2 \sigma$ error of the baseline noiseless case. 

We see that degrading the periods has the largest effect on $\Delta_{\rm rel}$, where the model is quite robust against a degradation in the amplitudes. As expected, we obtain the largest performance reduction, if all features are degraded. For ${\rm S/N} \approx 10^{3}$, we obtain consistent results with the noiseless case. 

This apparent difference in robustness between periods and amplitudes can be explained by noting that the sequences shown in the Petersen diagrams (Fig. \ref{fig:petersen_diagram}) can only be resolved, if the periods can be determined  accurately. As a result, we can assume that quite precise period determinations are needed to robustly separate the Wood sequences, which will be essential for accurate predictions. The correlation between amplitude and Wesenheit shown in Fig.~\ref{fig:features} are much broader and thus less sensitive to inaccurate measurements of the amplitudes. Greater sensitivity to  period measurements compared with measurements of amplitudes, can also be explained by the high feature importance of periods as discussed in \S \ref{app:feat_imp}.

This analysis considered the case of the same input feature noise across the training and test sets. We leave the case of different noise levels between training and test sets for future work. However we would like to note that the training sample will most likely have better photometry than the test sample, if nearby samples are used as anchors. This suggests that we can then artificially degrade this training sample to match the noise properties of the test sets, even without a more advanced correction. 
}

\bsp	
\label{lastpage}
\end{document}